\documentclass[11pt]{article}
\usepackage{amsmath}
\usepackage{amsfonts}
\usepackage{graphicx}
\usepackage[sort,compress]{cite}
\usepackage{lineno}
\usepackage{setspace}

\begin{document}
\begin{center}
\Large\bf{Conditions for successful data assimilation}
\end{center}

\begin{center}
Alexandre J. Chorin$^{a,b}$ and Matthias Morzfeld $^{a,b, }$\footnote{Email: mmo@math.lbl.gov}
\vspace{3mm}

$^a$Department of Mathematics\\
University of California, Berkeley;\\
\vspace{1mm}

$^b$Lawrence Berkeley National Laboratory.\\
\end{center}

\begin{center}
\emph{Abstract}
\end{center}
We show, using idealized models, that numerical data assimilation can be successful only if an 
effective dimension of the problem is not excessive. This effective dimension depends on the noise in the model and the data,
and in physically reasonable problems it can be moderate even when the number of variables is huge.
We then analyze several data assimilation algorithms, including particle filters and variational methods.
We show that well-designed particle filters can solve most of those data assimilation problems that can be solved in principle, and compare the conditions under which variational methods can succeed to the conditions required of particle filters. We also discuss the limitations of our analysis.

\section{Introduction}
Many applications in science and engineering require that the predictions of uncertain models be updated by information from a stream of noisy data (see e.g. \cite{Doucet2001,PeterJan2009,Bocquet2010}). The model and data jointly define a conditional probability density function~(pdf) $p(x^{0:n}|z^{1:n})$, where the discrete variable $n=0,1,2,\dots $ can be thought of as discrete time, $x^n$ is a real $m$-dimensional vector to be estimated, called the ``state", $x^{0:n}$ is a shorthand for the set of vectors $\{x^0,x^1,\dots,x^n\}$, and where the data sets 
$z^{n}$ are a $k$-dimensional vectors ($k\leq m$). All information about the state at time $n$ is contained in this conditional pdf 
and a variety of methods are available for its study, e.g. the Kalman filter~\cite{Kalman1960}, the extended and ensemble Kalman filter~
\cite{EvensenBook}, particle filters~\cite{Doucet2001}, or variational methods~\cite{TalagrandCourtier,Bennet1993}. Given a model and data, each of these algorithms will produce a result. We are interested in the conditions under which this result is reasonable, i.e. consistent with the real-life situation one is modeling. 

We say that data assimilation is feasible in principle, if it is possible to calculate the mean of the conditional probability density that it defines with a small-to-moderate uncertainty; we discuss what we mean
by ``moderate" below after we develop the appropriate tools. If data
assimilation is feasible in this sense, it is possible to find an estimate
of the state of a system whose distance from an outcome of the physical
experiment described by the dynamics is small-to-moderate,
with a high probability, i.e. reliable conclusions can be reached based on the results of the assimilation. We consider a data assimilation algorithm, e.g. a particle filter or a variational method, to be successful of it can produce an accurate estimate of the state of the system. A data assimilation algorithm can only be successful if data assimilation is feasible in principle. Our definition of success is in line with what is required in the physical sciences, where one wants to make reliable predictions given a model and data. We do not consider data assimilation to be successful if the posterior variance is reduced (e.g. when compared to the variance of the data) but remains large.

Generally, we restrict the analysis to linear state space models driven by Gaussian noise and supplemented by a synchronous stream of data perturbed by Gaussian noise, i.e. the noisy data are available at every time step of the model and only then. We further assume that all model parameters (including the covariance matrices of the noise) are known, i.e. we consider state estimation rather than combined state and parameter estimation. We study this class of problems because it can be examined in some generality and we can explain qualitatively its important aspects; however, we also discuss its limitations.

In section~2 we derive conditions under which data assimilation is feasible in principle, without regard to a specific algorithm. We define the effective dimension of a Gaussian data assimilation problem as the Frobenius norm of the steady state posterior covariance, and show that data assimilation is feasible in the sense described above only if this effective dimension is moderate. We argue that realistic problems have a moderate effective dimension. 

In the remainder of the paper we discuss the conditions under which particular data assimilation algorithms can succeed in solving problems (where success is defined as above) that are solvable in principle. In section 3 we briefly review particle filters. In section 4, we use the results of \cite{Snyder} to show that the optimal particle filter (which in the linear synchronous case coincides with the implicit particle filter \cite{atkins,chorin2010,Morzfeld2011}) performs well if the problem is solvable in principle, provided a certain balance condition is satisfied. We conclude that optimal particle filters can solve many data assimilation problems even if the number of variables to be estimated is large. Building on the results in\cite{Bickel,Bickel2,BickelBootstrap}, we show that another filter fails under conditions that are frequently met. Thus, how a particle filter is implemented is very important, since a poor choice of algorithm may lead to poor performance. In section 5 we consider particle smoothing and variational data assimilation and show that these methods as well can only be successful under conditions comparable to those we found in particle filtering. We discuss limitations of our analysis in section~\ref{sec:limitation} and present conclusions in section~7.

The effective dimension defined in the present paper is different from the effective dimensions introduced in \cite{Bickel,Bickel2,BickelBootstrap,Snyder}. The effective dimensions in\cite{Bickel,Bickel2,BickelBootstrap,Snyder} are defined for particular particle filters, whereas the effective dimension defined in the present paper is a characteristic of the model and data stream, i.e. independent of the data assimilation algorithm used. We show in particular that the effective dimension (as defined in the present paper) remains moderate for realistic models, even when the state dimension is large (asymptotically infinite), and that numerical data assimilation can be successful in these cases; in particular, a moderate effective dimension in our sense can imply moderate effective dimensions in the sense of \cite{Bickel,Bickel2,BickelBootstrap,Snyder} for a suitable algorithm.

\section{The effective dimension of linear Gaussian data assimilation problems}
\label{sec:EffectiveDimension}
We consider autonomous, linear Gaussian state space models of the form
\begin{linenomath*}
\begin{equation}
x^{n+1}=Ax^n+w^n
\label{model}
\end{equation}
\end{linenomath*}
where $n=0,1,2,\dots$ is a discrete time, $A$ is a given $m\times m$ matrix and $w^n$ are independent and identically distributed (iid) Gaussian random variables with mean zero and given covariance matrix $Q$, which we write 
as $w^n\sim\mathcal{N}(0,Q)$. The initial conditions may be random and we assume that their pdf is also Gaussian, i.e. $x^0\sim\mathcal{N}
(\mu_0,\Sigma_0)$, with both $\mu_0$ and $\Sigma_0$ given. We assume further that the data satisfy 
\begin{linenomath*}
\begin{equation}
z^{n+1}=Hx^{n+1}+v^{n+1},
\label{data}
\end{equation}
\end{linenomath*}
where $H$ is a given $k\times m$ matrix ($k\leq m$) and the $v^{n+1}\sim\mathcal{N}(0,R)$ are iid, where $R$ is a given $k\times k$ matrix. The $w^n$'s and $v^n$'s are independent of each other and also independent of $x_0$. 

In principle, but not necessarily in practice, the
covariance matrix $P_n$ of the state $x^n$ conditioned on the data $z^{1:n}$ can be computed recursively, starting with $P_0 = \Sigma_0$:
\begin{linenomath*}\begin{eqnarray*}
X_n &=&AP_nA^T+Q,\\
K_n& =& X_nH^T(HX_nH^T+R)^{-1},\\
P_{n+1} &=& (I_m-K_nH)X_n,
\end{eqnarray*}\end{linenomath*}
where $I_m$ is the identity matrix of order $m$ and the $m\times k$ matrix $K_n$ is often called the ``Kalman gain''. This is the Kalman formalism. We assume that the pair $(H,A)$ is $d$-detectable and that $(A,Q)$ is $d$-stabilizable. Detectability and stabilizabilty can respectively be interpreted (roughly) as requiring that the observation operator be sufficiently rich to determine the dynamics and the noise be able to affect the whole dynamics (see \cite{Lancaster}, pp. 90--91 for technical definitions). These assumptions allow unstable dynamics, as often encountered in geophysics, but also make it possible to perform a steady state analysis because the covariance matrix reaches a steady state so that 
\begin{linenomath*}\begin{equation*}
P_{n+1} =P_n = P = (I-KH)X,
\end{equation*}\end{linenomath*}
where $X$ is the unique positive semi-definite solution of the discrete algebraic Riccati equation (DARE)
\begin{linenomath*}\begin{equation*}
X=AXA^T-AXH^T(HXH^T+R)^{-1}HXA^T+Q,
\end{equation*}\end{linenomath*}
and where 
\begin{linenomath*}\begin{equation*}
K = XH^T(HXH^T+R)^{-1},
\end{equation*}\end{linenomath*}
is the ``steady state'' Kalman gain. Note that the steady state covariance matrix $P$ is independent of the initial covariance matrix $\Sigma_0$ and that the rate of convergence to this limit is at least linear, in many cases quadratic (see \cite{Lancaster}, p. 313). This means that, after a relatively short time, the samples of the state given the data are normally distributed with mean $\mu_n$ and covariance matrix $P$ (the mean $\mu_n$ of the variables is not needed here, but it can also be computed using Kalman's formulas). 

The steady state covariance matrix, $P=(p_{ij})$ determines the posterior uncertainty, i.e. the uncertainty after we considered the data. If $P$ is ``large'', the uncertainty is large, which translates to a large spread of the samples in state space. We suggest to measure uncertainty with the Frobenius norm of $||P||_F=(\sum_{ij}p_{ij}^2)^{1/2}$, because this norm determines the spread of the posterior samples in state space. 

To see this, consider the random variable $y = (x_n-\mu_n)^T(x_n-\mu_n)$, where $x_n-\mu_n\sim\mathcal{N}(0,P)$, i.e. consider the squared distances of the samples from their mean (their most likely value). 
Let $U$ be an orthogonal $m\times m$ matrix whose columns are the eigenvectors of $P$. Then
\begin{linenomath*}\begin{equation*}
y= (x_n-\mu_n)^T(x_n-\mu_n)=s^Ts=\displaystyle \sum_{j=1}^ms_j^2,
\end{equation*}\end{linenomath*}
where $s=U^T(x_n-\mu_n)\sim\mathcal{N}(0,\Lambda)$, and $\Lambda=U^TPU$ is a diagonal matrix whose diagonal elements are the $m$ 
eigenvalues $\lambda_j$ of $P$. It is now straightforward to compute the mean and variance of $y$ because the $s_j$'s (the elements of $s$) are independent:
\begin{linenomath*}\begin{equation*}
E(y) = \displaystyle \sum_{j=1}^m \lambda_j,\quad var(y) = \displaystyle 2\sum_{j=1}^m \lambda_j^2.
\end{equation*}\end{linenomath*}
Note that $y=r^2$, where $r$ is the distance from the sample to the most likely state (the mean). Assuming that $m$ is large, we obtain, using Taylor expansion of  $r/\sqrt{\sum \lambda_j}=(y/\sum \lambda_j)^{1/2}$ around $1$ and assuming that $\lambda_j=O(1)$, that
\begin{linenomath*}
\begin{eqnarray*}
E(r)  &=& \frac{2\left(\displaystyle\sum_{j=1}^m \lambda_j\right)^2-\displaystyle\sum_{j=1}^m \lambda_j^2}{2\left(\displaystyle\sum_{j=1}^m \lambda_j\right)^{1.5}}+O_p\left(\frac{\displaystyle\sum_{j=1}^m \lambda_j^4}{\left(\displaystyle\sum_{j=1}^m \lambda_j\right)^4}\right)=\hat{E}(r)+O_p\left(\frac{\displaystyle\sum_{j=1}^m \lambda_j^4}{\left(\displaystyle\sum_{j=1}^m \lambda_j\right)^4}\right),\\
var(r )&=&\frac{\displaystyle \sum_{j=1}^m \lambda_j^2}{2\displaystyle \sum_{j=1}^m \lambda_j}+O_p\left(\frac{\displaystyle\sum_{j=1}^m \lambda_j^4}{\left(\displaystyle\sum_{j=1}^m\lambda_j\right)^3}\right)=\hat{v}(r)+O_p\left(\frac{\displaystyle\sum_{j=1}^m \lambda_j^4}{\left(\displaystyle\sum_{j=1}^m\lambda_j\right)^3}\right).
\end{eqnarray*}
\end{linenomath*}
The techniques in \cite{BickelBootstrap} can be used to extend the above formulas  for $m\rightarrow \infty$, $\sum\lambda \rightarrow \infty$ and with $\lambda_j=O(1)$, i.e. to the case for which the moments of $y$ do not necessarily exist. We use standard inequalities to show that
\begin{linenomath*}\begin{equation*}
	\sqrt{\displaystyle \sum_{j=1}^m \lambda_j^2} \leq \displaystyle \sum_{j=1}^m \lambda_j \leq \, \sqrt{m\displaystyle \sum_{j=1}^m \lambda_j^2},
\end{equation*}\end{linenomath*}
and, with these, obtain bounds for $\hat{E}$ and $\hat{v}$:
\begin{linenomath*}\begin{equation*}
m^{3/4}\left(\displaystyle\sum_{j=1}^m \lambda_j^2\right)^{1/4} \leq \hat{E} \leq  m\left(\displaystyle\sum_{j=1}^m \lambda_j^2\right)^{1/4}, \quad \frac{1}{2\sqrt{m}}\left(\displaystyle\sum_{j=1}^m \lambda_j^2\right)^{1/2}\leq\hat{v}\leq \frac{1}{2}\left(\displaystyle\sum_{j=1}^m \lambda_j^2\right)^{1/2}.
\end{equation*}\end{linenomath*}
The Frobenius norm of a matrix is the square root of the sum of its eigenvalues squared, i.e. $||P||_F =\sqrt{\sum\lambda^2}$. Thus, the above  bounds indicate that the Frobenius norm of $P$ determines the mean and variance of the distance of a sample from the most likely state, i.e. the spread of the samples in the state space.  

Based on the calculations above, we now investigate what a large posterior covariance, i.e. a large spread of posterior samples, means for data assimilation. Suppose that $m$ is large and that $\lambda_j =O(1)$ for $j=1,\dots,m$; then $\hat{E} =O(m^{1/2})$ and $\hat{v} =O(1)$. This means that the samples collect on a shell of thickness $O(1)$ at a distance $O(m^{1/2})$ from their mean and are distributed over a volume $O(m^{(m+1)/2})$, i.e., for large $m$, the predictions spread out over a large volume at a large distance from the most likely state. By considering both the model (\ref{model}) and the data (\ref{data}), one concludes that the true state is likely to be found somewhere on this shell. However, since this shell is huge, the various states on it can correspond to very different physical situations. Knowing that the state is somewhere on this shell is not satisfactory if one wants to compute a reliable estimate of the state; the uncertainties in the model and the observation error are too large. 

What we have shown is that data assimilation makes sense, according to our definitions, only if the Frobenius norm of the posterior steady state covariance matrix is moderate. We thus define the effective dimension of the Gaussian data assimilation problem defined by equations~(\ref{model}) and~(\ref{data}) to be this Frobenius norm:
\begin{linenomath*}\begin{equation*}
	m_{\text{eff}}\doteq  ||P ||_F = \sqrt{\sum_{j=1}^m \lambda_j^2}. 
\end{equation*}\end{linenomath*}
Data assimilation can only be successful if this effective dimension is moderate. The precise value of the effective dimension that can not be exceeded if one wants to reach reliable conclusions varies from one problem to the next and, in particular, depends on the level of accuracy required, so that it is very difficult to pin down an upper bound for the effective dimension in general. In cases where one can interpret the data assimilation problem defined by (\ref{model}) and (\ref{data}) as an approximation to an infinite dimensional problem, e.g. in problems that arise from partial differential equations (PDE), our requirements imply that the effective dimension remains bounded as $m\rightarrow \infty$. This is connected to well-posedness, stability and accuracy of infinite dimensional Bayesian inverse problems discussed in \cite{Stuart}. 

We expect that the effective dimension is moderate in practice, since the data assimilation problem reflects an experimental situation, and we wish that the numerical samples behave like experimental samples: if the uncertainty is large, one will observe that the outcomes of repeated experiments exhibit a large spread; if the uncertainty is small, then the spread in the outcomes of experiments is also small. Since the outcomes of repeated experiments rarely exhibit large variations, one should expect that the samples of numerical data assimilation all fall into a small ``low-dimensional" ball, centered around the most likely state, i.e. the radius, $E( r)\approx\hat{E}$, is comparable to the thickness, $var( r)\approx\hat{v}$ (see below). 

For the reminder of this section we will investigate conditions for successful data assimilation by studying conditions on the errors in the model (\ref{model}), represented by the covariance matrix~$Q$, and conditions on the errors in the data (\ref{data}), represented by the covariance matrix $R$, that lead to a moderate effective dimension.

Finally, we point out that the effective dimension defined above is different from the effective dimensions defined in \cite{Bickel,Bickel2,BickelBootstrap,Snyder}, which came up in connection with specific particle filters. The effective dimension defined here is defined from the posterior pdf and, thus, is independent of a data assimilation technique; it is a characteristic of the model~(\ref{model}) and data stream~(\ref{data}). However, since we consider the posterior pdf of linear Gaussian data assimilation problems (for which the Kalman formalism gives the answer), our analysis is valid only for such models. We discuss the limitations of our analysis in more detail in section \ref{sec:limitation}.

\subsection{Bounds on the effective dimension}\label{sec:EffDimBounds}
To discover the real-life interpretation of the effective dimension, we study its upper bounds in terms of the Frobenius norms of $Q$ and $R$.  From Khinchin's theorem (see e.g. \cite{ChorinHald}) we know that the Frobenius norms of $Q$ and $R$ must be bounded as $m,k\rightarrow\infty$ or else the energies of the noises are infinite, which is unrealistic. We show that a moderate Frobenius norm of $Q$ and $R$ can lead to a moderate effective dimension. We start by a simple example, which is also useful in the study of data assimilation methods in later sections.

\vspace{3mm}

\subsubsection{Example} Put $A=H=I_m$ and let $Q=qI_m$, $R=rI_m$. The Riccati equation can be solved analytically for this example and we find the effective dimension
\begin{linenomath*}\begin{equation*}
m_{\text{eff}}=\sqrt{m}\frac{\sqrt{q^2+4qr}-q}{2}.
\end{equation*}\end{linenomath*}
In a real-life problem, we would expect $||P||_F$ and thus $m_{\text{eff}}$ to grow slowly, if at all, when the number of variables increases. In fact, we have just shown that $m_{\text{eff}}$ must be moderate or else data assimilation can not be successful.

The condition of moderate effective dimension induces a ``balance condition'' between the errors in the model (represented by $q$) and the errors in the data (represented by~$r$). In this simple example, an $O(1)$ effective dimension gives rise to the balance condition
\begin{linenomath*}\begin{equation*}
\frac{\sqrt{q^2+4qr}-q}{2}\leq \frac{1}{\sqrt{m}},
\end{equation*}\end{linenomath*}
where the $1$ in the numerator of the right-hand side stands for a constant; we set this constant equal to $1$ because this already captures the general behavior. The constant cannot be pinned down precisely because an acceptable level of accuracy may vary from one application to the next; the balance condition above, and its generalizations below, do however provide guidance as to what can be done. 

\begin{figure}[h!tbp]
\begin{center}
{\includegraphics[width=.7\textwidth]{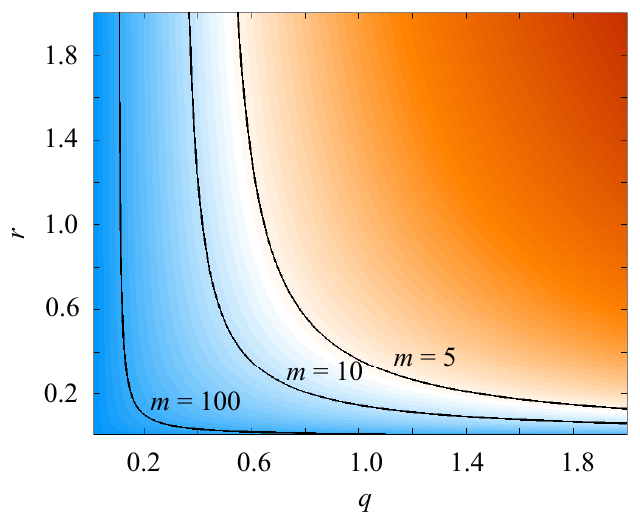}}
\caption{Conditions for successful sequential data assimilation.} 
\label{fig:KFMap} 
\end{center}
\end{figure}

Figure \ref{fig:KFMap} illustrates the condition for successful data assimilation and shows a plot of the function that is defined by the left-hand-side of the above inequality as well as three level sets, corresponding to $m=5,10,100$ respectively; for a given dimension~$m$, all values of~$q$ and~$r$ below the corresponding level set lead to an $O(1)$ effective dimension, i.e. to a scenario in which data assimilation is feasible in principle.

The condition implies that, for fixed $m$, the smaller the errors in the data (represented by $r$), the larger can be the uncertainty in the model (represented by $q$) and vice versa. Moreover, note that for very small $q$, the boundaries for successful data assimilation are (almost) vertical lines. The reason is that if the model is very good, neither accurate nor inaccurate data can improve it, i.e. data assimilation is not necessary. If the model is poor, only nearly perfect data can help. We will encounter this balance condition (in more complicated forms) again in the general case in the next section and also in the analysis of particle filters and variational data assimilation.

Finally, note that the Frobenius norms $||Q||_F=q\sqrt{m}$ and $||R||_F=r\sqrt{m}$ increase with the number of dimensions unless $q$ or $r$ or both decrease with $m$ as shown in figure \ref{fig:KFMap}. We will argue in section \ref{sec:EffDimRealLife} that in realistic cases, the Frobenius norms of $Q$ and $R$ are moderate even if $m$ or $k$ are large (asymptotically infinite). We also expect, but cannot prove in general, that a balance condition as in figure \ref{fig:KFMap} is valid in the general case (arbitrary $A,H,Q,R$), with $q$ and $r$ replaced by the Frobenius norms of $Q$ and $R$.
\vspace{3mm}

\subsubsection{The general case}
In the general case, the condition for successful data assimilation that must be satisfied by uncertainties in the model ($||Q||_F$) and data ($||R||_F$) is more complicated because the effective dimension is the Frobenius norm of the solution of a Riccati equation which in general does not admit a closed form solution. 

However, if the covariance matrices $Q$ and $R$ have moderate Frobenius norms, then the effective dimension of the problem can be moderate even if $m$ and $k$ are large and, thus, data assimilation can be successful. To see this, let $X$ and $P$ be the solution of the DARE respectively the steady state covariance matrix of a given $(A,Q,H,R)$ data assimilation problem and let $\tilde Q\leq Q$, i.e. $\tilde Q - Q$ is symmetric positive semi-definite (SPD). If $\tilde R\leq R$, then, by the comparison theorem (Theorem 13.3.1) in \cite{Lancaster}, $\tilde X \leq X$, where $\tilde X$ is the solution of the DARE associated with the $(A,\tilde Q,H,\tilde R)$ data assimilation problem. From the Kalman formulas we know that
\begin{linenomath*}\begin{equation*}
P = X-XH^T(HXH^T+R)^{-1}HX,
\end{equation*}\end{linenomath*}
which implies that $P\leq X$. Moreover, for two SPD matrices $C$ and $D$, $C\leq D$ implies $||C||_F \leq ||D||_F$. Thus, the smaller the Frobenius norm of $Q$ and $R$, the smaller is the upper bound $||X||_F$ on the effective dimension. 

However, the requirement that these Frobenius norms be moderate is not sufficient to ensure that the effective dimension of the problem is moderate; in particular, it is evident that the properties of $A$ must play a role; for example, if the $L_2$ norm of $A$ exceeds unity, the model (\ref{model}) is unstable and successful data assimilation is unlikely unless the data are sufficiently rich to compensate for the instabilities (see also \cite{Stuart}). We have assumed such difficulties away by assuming the pair $(H,A)$ to be $d$-detectable and $(A,Q)$ to be $d$-stabilizable. However, unstable dynamics should be treated carefully and in specific cases (for nonlinear problems) as in \cite{Brett2013}.

While the model, or $A$, is implicitly accounted for in $X$, the solution of the DARE, one can construct sharper bounds on the effective dimension by accounting for the model~(\ref{model}) and data stream~(\ref{data}) more explicitly. To that extent, we construct matrix bounds on $P$, from matrix bounds for the solution of the DARE \cite{Kwon}. Let $X\leq X_u$, and $X_l\leq X$, be upper and lower matrix bounds for the solution of the DARE, for example, we can choose the lower bound in \cite{Komaroff1994}
\begin{linenomath*}\begin{equation*}
 Q\leq X_l = A(Q^{-1}+H^TR^{-1}H)^{-1}A^T+Q \leq X,
\end{equation*}\end{linenomath*}
and the upper bound in \cite{Kwon}
\begin{linenomath*}\begin{equation*}
  X \leq X_u = A(X_*^{-1}+H^TR^{-1}H)^{-1}A^T+Q,
\end{equation*}\end{linenomath*}
where $X_* = A(\eta^{-1}+H^TR^{-1}H)^{-1}A^T+Q$, $\eta =f(-\lambda_1(A)-\lambda_n(H^TR^{-1}H)\lambda_1(Q)+1,2\lambda_n(H^TR^{-1}H),2\lambda_1(Q)))$, $f(a,b,c) = (\sqrt{a^2+bc}-a)/2)$ and $\lambda_1( C)$ and $\lambda_n( C)$ are the largest respectively smallest eigenvalue of the matrix $C$. Then an upper matrix bound for the steady state covariance matrix is
\begin{linenomath*}\begin{equation*}
P \leq X_u-X_lH^T(HX_uH^T+R)^{-1}HX_l.
\end{equation*}\end{linenomath*}
The Frobenius norm of this upper matrix bound is an upper bound for the effective dimension.
\vspace{3mm}

\subsection{The real-world interpretation of effective dimension}\label{sec:EffDimRealLife}
We have shown that there is little hope for reaching reliable conclusions unless the effective dimension of the data assimilation problem defined by equations~(\ref{model}) and~(\ref{data}) is moderate. We now give more detail about the physical interpretation of this result.  

Suppose the variables $x$ one is estimating are point values of, for example, the velocity of a flow field (as they often are in 
applications). The Frobenius norm of the covariance matrix $Q$ is proportional to the specific kinetic energy of the noise field that is perturbing an underlying flow. This energy should be a small fraction of the energy of the flow, or else there is not enough information in the model (\ref{model}) to examine the flow one is interested in. We can thus assume that the Frobenius norm of $Q$ is moderate. By the same arguments, we can assume that the Frobenius norm of $R$ is moderate, or else the noise in the data equation overpowers the actual measurements. Since moderate Frobenius norms of $Q$ and $R$ often imply a moderate Frobenius norm of $P$, we typically are dealing with a data assimilation problem with a moderate effective dimension, even if $m$ and $k$ are arbitrarily large.  

Point values of a flow field usually come from a discretization of a stochastic differential equation. As one refines this discretization, one can expect the correlation between the errors at neighboring grid-points to increase. These errors are represented by the covariance matrix $Q$ and from Khinchin's theorem (see e.g. \cite{ChorinHald}) we know that a random field with sufficiently correlated components has a finite energy density (and vice versa). This implies for the finite dimensional case that the Frobenius norm of $Q$ does not grow without bound as we increase $m$.

Another and perhaps even more dramatic instance of this situation is one where the random process we are interested in is smooth so that the spectrum of its covariance matrix decays quickly \cite{Adler,Rasmussen}. For practical purposes one may then consider $m-d$ of the eigenvalues to be equal to zero (rather than just very small). This is an instance of ``partial noise'' \cite{Morzfeld2012}, i.e. the state 
space splits into two disjoint subspaces, one of dimension $d$, which contains state variables, $u$, that are directly driven by Gaussian 
noise, and one of dimension $m-d$, which contains the remaining variables, $v$, that are (linear) functions of the random variables~$u$. Thus, the steady state covariance matrix is of size $d\times d$ and the effective dimension is independent of the state dimension and moderate even if $m$ is large. Smoothness of the random perturbations may be particularly important in data assimilation for PDE (e.g. in fluid mechanics), since the PDE itself can require regularity conditions \cite{Stuart}.

Note that the key to the moderate effective dimension in all of the above cases is the correlation among the errors and indeed, the data assimilation problems derived by various practitioners and theorists show a strong correlation of the errors (see e.g. \cite{vanLeuween2003,Wheeler,Zhang,Rasmussen,Adler,MillerCane,MillerHackert,MillerSpitz,Morzfeld2012,Bennet1987}). The correlations are also key to the well-boundedness of infinite dimensional problems \cite{Stuart} where the spectra of the covariances (which are compact operators in this case) decay; a well correlated noise model was obtained from an infinite dimensional problem in \cite{Bennet1987}.

The geometrical interpretation of this situation is as follows: because of correlations in the noise, the probability mass is concentrated on a $d$-dimensional manifold, regardless of the dimension $m\geq d$ of the state space. In addition one must be careful that the noise in the 
observations not be too strong. Otherwise the data can push the probability mass away from the $d$-dimensional manifold (i.e. the data 
increase uncertainty, instead of decreasing it). This assumption is reasonable, because typically the data contain information and not just noise. Similar observations were reported for infinite dimensional, strong constraint problems for low-observation noise (covariance of the error in the data goes to $0$), see Theorem 2.5 in \cite{Stuart}. 

Next, suppose that the vector $x$ in (\ref{model}) and (\ref{data}) represents the components of an abstract model with the several components representing various indicators, for example of economic activity (so that the concept of energy is not well-defined). It is unreasonable to assume that each source of error affects only one component of $x$. As an example of what happens when each source of error affects many components, consider a model where Gaussian sources of error are distributed with spherical symmetry in the space of the $x$'s and have a magnitude independent of the dimension $m$. In an $m$ dimensional space, the components of the unit vector of length $1$ have magnitude of order $O(m^{-0.5})$, so that the variance of each component must decrease like $m^{-1}$. Thus, the covariance matrices in~(\ref{model}) and~(\ref{data}) are proportional to $m^{-1}I_m$ and the effective dimension (for $A=H=I_m$) is $||P||_F=(\sqrt{5}-1)/2m$, which is small when $m$ is large. This is a plausible outcome, because the more data and indicators are considered, the less uncertainty there should be in the outcome (because the new indicators provide additional information).

\section{Review of particle filters}
\label{sec:Review}
In importance sampling one generates samples from a hard-to-sample pdf $p$ (the ``target" pdf) by producing weighted samples from an easy-to-sample pdf, $\pi$, called the ``importance function" (see e.g. \cite{KalosWhitlock,ChorinHald}). Specifically, if the random variable one is interested in is $x \sim p$, one generates samples $X_j \sim \pi, j=1,\dots,M,$ (we 
use capital letters for realizations of random variables) and weighs each by the weight 
\begin{linenomath*}\begin{equation*}
	W_j \propto \frac{p(X_j)}{\pi(X_j)}.
\end{equation*}\end{linenomath*}
The weighted samples $\left\{ X_j,W_j \right\}$ (called particles in this context) form an empirical estimate of the target pdf $p$, i.e. for a smooth function $u$, the sum 
\begin{linenomath*}\begin{equation*}
	E_M(u) = \sum_{j=0}^{M}u(X_j)\hat{W}_j,
\end{equation*}\end{linenomath*}
where $\hat{W}_j=W_j/\sum_{j=0}^{M}W_j$, converges almost surely to the expected value of $u$ with respect to the pdf $p$ as $M \rightarrow 
\infty$, provided that the support of $\pi$ includes the support of~$p$. 

Particle filters apply these ideas to the recursive formulation of the conditional pdf:
\begin{linenomath*}\begin{equation*}
	p(x^{0:n+1}|z^{1:n+1})=p(x^{0:n}|z^{1:n})\frac{p(x^{n+1}|x^{n})p(z^{n+1}|x^{n+1})}{p(z^{n+1}|z^{1:n})}.
\end{equation*}\end{linenomath*}
This requires that the importance function factorize in the form:
\begin{linenomath*}
\begin{equation}
\label{eq:factorization}
\pi(x^{0:n+1}|z^{0:n+1}) = \pi_0(x^0)\prod_{k=1}^{n+1} \pi_k(x^{k}|x^{0:k-1},z^{1:k}).
\end{equation}
\end{linenomath*}
where the $\pi_k$ are updates for the importance function.
The factorization of the importance function leads to the recursion 
\begin{linenomath*}
\begin{equation}
\label{eq:Weights}
 W^{n+1}_j\propto \hat{W}^{n}_j\frac{p(X_j^{n+1}|X_j^{n})p(Z^{n+1}|X_j^{n+1})}{\pi_{n+1}(X_j^{n+1}|X_j^{0:n},Z^{0:k})},
\end{equation}
\end{linenomath*}
for the weights of each of the particles, which are then scaled so that their sum equals one. Using ``resampling'' techniques, i.e. replacing particles with small weights with ones with large weights (see e.g. \cite{Doucet2001,GordonSIR} for resampling algorithms), makes it possible to set $\hat{W}_j^n=1/M$ when one computes $W^{n+1}_j$. Once one has set $\hat{W}_j^n=1/M$ but before sampling a new state at time $n+1$, each of the weights can be viewed as a function of the
random variable $x^{n+1}_j$ and is therefore a random variable. 

The weights determine the efficiency of particle filters. Suppose that, before the normalization and resampling step, one weight is much larger than all others; then upon rescaling of the weights such that their sum equals one, one finds that the largest normalized weight is near 1 and all others are near 0. In this case the empirical estimate of the conditional pdf by the particles is very poor (it is a single, often unlikely point) and the particle filter is said to have collapsed. The collapse of particle filters can be studied via the variance of the logarithm of the weights, and it was argued rigorously in \cite{Bickel,Bickel2,BickelBootstrap,Snyder} that a large variance of the logarithm of the weights leads to the collapse of particle filters. The choice of importance function $\pi$ is critical for avoiding the collapse and many different importance functions have been considered in the literature (see e.g. \cite{Brad ,Weare2009,Weare2012,vanLeeuwen,Ades,Chorintupnas,chorin2010,Morzfeld2011}). Here we we follow \cite{Bickel,Bickel2,BickelBootstrap,Snyder} and discuss two particle filters in detail.

\subsection{The SIR filter}
A natural choice for the importance function is to generate samples with the model (\ref{model}), i.e. to choose $\pi_{n+1}=p(x^{n+1}|x^n)$. When a resampling step is added, the resulting filter is often called a sequential importance sampling with resampling (SIR) filter\cite{GordonSIR} and its weights are
\begin{linenomath*}\begin{equation*}
 W^{n+1}_j\propto p(Z^{n+1}|X_j^{n+1}).
 \end{equation*}\end{linenomath*}
It is known that the SIR filter collapses if the probability measure induced by the importance function $\pi_{n+1}=p(x^{n+1}|x^n)$, and the probability measure induced by the target pdf, $p(y^{n+1}|x^{n+1}) p(x^{n+1}|x^n)$, have supports such that an event that has significant probability in one of them has a very small probability in the other. This can happen even in one dimensional problems, however the situation becomes more dramatic as the dimension $m$ increases. A rigorous analysis of the asymptotic behavior of weights of the SIR filter (as the number of particles and the dimension go to infinity) is given in \cite{Bickel,Bickel2,BickelBootstrap} and it is shown that the number of particles required to avoid the collapse of the SIR filter grows exponentially with the variance of the observation log likelihood (the logarithm of the weights).

\subsection{The optimal particle filter}\label{sec:OptimalFilter}
One can avoid the collapse of particle filters in low-dimensional problems by choosing the importance function wisely. If one chooses an importance function $\pi$ so that the weights in (\ref{eq:Weights}) are close to uniform, then all particles contribute equally to the empirical estimate they define. In \cite{Doucet,OptimalImportanceFunction,liuchen1995,Snyder} the importance function
$\pi_{n+1}(x^{n+1}|x^{0:n},z^{0:n+1})= p(x^{n+1}|x^{n},z^{n+1})$, is discussed and it is shown that this importance function is``optimal'' in the sense that it minimizes the variance of the weights given the data and $X_j^n$. For that reason, a filter that uses this importance function is called ``optimal particle filter'' and the optimal weights are
\begin{linenomath*}
\begin{equation*}
W_j^{n+1} \propto p(Z^{n+1}|X_j^{n}).
\end{equation*}
\end{linenomath*}
For the class of models and data we consider, the optimal particle filter is identical to the implicit particle filter \cite{atkins,Morzfeld2011,chorin2010}. The asymptotic behavior of the weights of the optimal particle filter was studied in \cite{Snyder} and it was found that the optimal filter collapses if the variance of the logarithm of its weights is large. A connection to the collapse of the implicit particle filter (for linear Gaussian models) was made in \cite{Ades}.

\section{The collapse and non-collapse of particle filters}
The conditions for the collapse have been reported in \cite{Bickel,Bickel2,BickelBootstrap} for SIR and in \cite{Snyder} for the optimal particle filter; here we connect these to our analysis of effective dimension.

\subsection{The case of the optimal particle filter}\label{sec:OptimalFilter}
It was shown in \cite{Snyder}, that the optimal particle filter collapses if the Frobenius norm of the covariance matrix of $\left(HQH^T+R\right)^{-0.5}HAx^{n-1}$ is large (asymptotically infinite as $k\rightarrow \infty$). However if this Frobenius norm is moderate, then the variance of the logarithm of the weights is also moderate so that the optimal particle filter works just fine (i.e. it does not collapse) even if $k$ is large. We now investigate the role the effective dimension of section~\ref{sec:EffectiveDimension} plays for the collapse of the optimal particle filter.

Following \cite{Snyder} and assuming that the conditional pdf has reached steady state, i.e. that the covariance of $x^{n-1}$ is $P$, the steady state solution of the Riccati equation, one finds that the Frobenius norm of the symmetric matrix
\begin{linenomath*}
\begin{equation}
\label{eq:EVP}
	\Sigma= HAP A^TH^T\left(HQH^T+R\right)^{-1},
\end{equation}
\end{linenomath*}
governs the collapse of the optimal particle filter. If the Frobenius norm of $\Sigma$ is moderate then the optimal particle filter will work, even for large~$m$ and~$k$. A condition for successful data assimilation with the optimal particle filter is thus that the Frobenius norm of $\Sigma$ is moderate. This condition induces a balance condition between the errors in the model and in the data, which must be satisfied or else the optimal particle filter will fail; the situation is analogous to what we observed in section \ref{sec:EffectiveDimension}. 

\begin{figure}[h!tbp]
\begin{center}
{\includegraphics[width=1.0\textwidth]{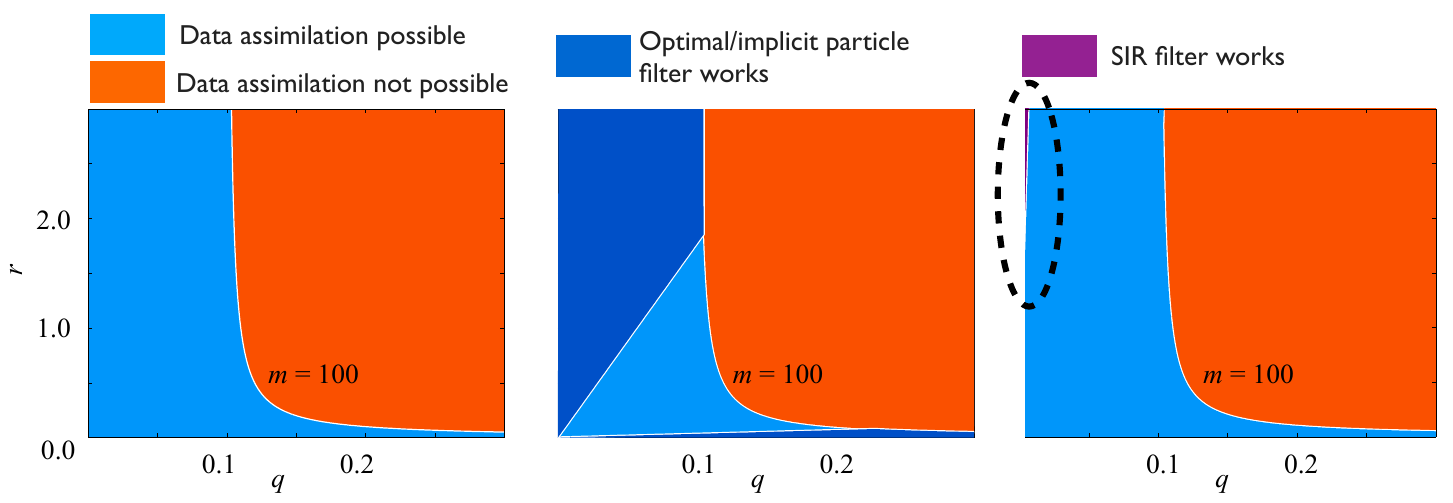}}
\caption{Conditions for successful sequential data assimilation (left panel), and for successful particle filtering; center panel: optimal\slash implicit particle filter; right panel: SIR filter. The broken ellipse in the right panel locates the area where the SIR filter works.} 
\label{fig:CompareMap} 
\end{center}
\end{figure}

To understand the balance condition better, we consider again the simple example of section \ref{sec:EffectiveDimension}, i.e. we set $H=A=I_m$ and $Q=qI_m$, $R=rI_m$. We already computed $P$ in section~\ref{sec:EffectiveDimension} and find from~(\ref{eq:EVP}) that
\begin{linenomath*}\begin{equation*}
||\Sigma||_F = \sqrt{m}\frac{\sqrt{q^2+4qr}-q}{2(q+r)}.
\end{equation*}\end{linenomath*}
so that the balance condition becomes
\begin{linenomath*}\begin{equation*}
\frac{\sqrt{q^2+4qr}-q}{2(q+r)}\leq \frac{1}{\sqrt{m}},
\end{equation*}\end{linenomath*}
where the $1$ in the numerator again stands for a constant $O(1)$, which we set equal to $1$ because this already captures the general behavior. Note that, for $m$ fixed, the left-hand-side depends only on the ratio of the covariances of the noise in the model and in the data, so that the level sets are rays. In the center panel of figure \ref{fig:CompareMap}, we superpose these rays, for which optimal particle filtering can be successful, with the ($q,r$)-region in which data assimilation is feasible in principle (as computed in section \ref{sec:EffectiveDimension}). The left panel of the figure shows what is in principle possible, for comparison.

We find that the optimal particle filter can successfully solve most of the data assimilation problems that are feasible to solve in principle (see section~\ref{sec:EffectiveDimension}). The exception are problems for which $q\approx r$, i.e. the noise in the model and data are equally strong. 

\begin{figure}[h!tbp]
\begin{center}
{\includegraphics[width=0.6\textwidth]{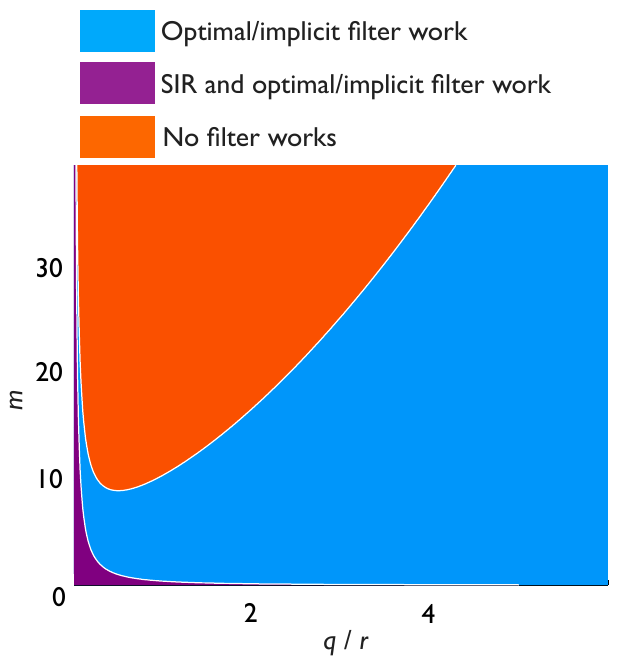}}
\caption{Maximum dimension for two particle filters.} 
\label{fig:FilterMap} 
\end{center}
\end{figure}

Another way to see this is to set $\epsilon =q/r$ so that the balance condition for successful optimal particle filtering becomes
\begin{linenomath*}\begin{equation*}
\frac{\sqrt{\epsilon^2+4\epsilon}-\epsilon}{2(1+\epsilon)}\leq \frac{1}{\sqrt{m}},
\end{equation*}\end{linenomath*}
which we solve for $m$ and then plot the maximum dimension $m$ as a function of the ratio of the noise in the model and the noise in the data; all values smaller than this maximum dimension are shown in figure \ref{fig:FilterMap} as the light blue area.
We conclude that the optimal particle filter works for high-dimensional data assimilation problems if $\epsilon$ is either small or large.  The case of large $\epsilon$ is the case typically encountered in practice. The reasons are as follows: if $\epsilon$ is small, then the model is very accurate. In this case, neither accurate nor inaccurate data can improve the model predictions (this case corresponds to the vertical line in figure \ref{fig:CompareMap}), i.e. data assimilation is unnecessary since one can simply trust the predictions of the model (\ref{model}). If $\epsilon$ is large, then the uncertainty in the data is much less than the uncertainty in the model, i.e. we can learn a lot from the data. This is the interesting case and the optimal particle filter (or the implicit particle filter) can be expected to work in such scenarios. However, problems occur when $\epsilon\approx 1$. We expect this case to occur infrequently, because typically the data are more accurate than the model. 

It is however important to realize that the collapse of the optimal particle filter for $\epsilon \approx 1$ does not imply that Monte Carlo sampling in general is not applicable in this case. Particle filtering induces variance into the weights because of its recursive problem formulation and this variance can be reduced by particle smoothing. The reason is as follows:  the variance of the weights of the optimal particle filter depends only on the variance of the particles' positions at time $n$ (see section \ref{sec:OptimalFilter}), i.e. each particle is updated to time $n+1$ such that no additional variance is introduced (this is why this filter is called optimal); however the particles at time $n$ may be unlikely in view of the data at $n+1$ (due to accumulation of errors up until this point). In this case, one can go back and correct the past, i.e. use a particle smoother (see also section~\ref{sec:VarAndSmoothing}). However, the number of steps one needs to go back in time for successful smoothing is problem dependent and, thus, we cannot provide a full analysis here (given that we work in a restrictive linear setting it seems more realistic to do this analysis on a case by case basis). In particular, it was indicated in two independent papers \cite{Weare2012,Brad} that smoothing a few steps backwards can help with making Monte Carlo sampling applicable in situations where particle filters fail or perform poorly. In \cite{Weare2012}, particle smoothing for the ``low-noise regime'' (which is an instance of the case where $\epsilon\approx 1$) is considered in connection with an application in oceanography. In~\cite{Brad}, particle smoothing was found to give superior results than particle filtering for combined parameter and state estimation, again in connection with an application in oceanography. However the approximations for (optimal) particle smoothers become difficult and computationally expensive as the problems get nonlinear.

In the general case (arbitrary $A,H,Q,R$), we can simplify the balance condition for successful particle filtering by using the upper bound for the Frobenius norm of $\Sigma$ :
\begin{linenomath*}\begin{equation*}
	||\Sigma||_F\leq ||A||_F^2 ||H||_F^2 ||P||_F ||\left(HQH^T+R\right)^{-1}||_F.
\end{equation*}\end{linenomath*}
If we require that this upper bound is less than $\sqrt{m}$, then we find, using the upper bound 
\begin{linenomath*}\begin{equation*}
\sqrt{m}=||I||_F \leq ||\left(HQH^T+R\right)||_F ||\left(HQH^T+R\right)^{-1}||_F,
\end{equation*}\end{linenomath*}
that
\begin{linenomath*}\begin{equation*}
	||A||_F^2 ||H||_F^2 ||P||_F \leq ||H||_F^2||Q||_F+||R||_F,
\end{equation*}\end{linenomath*}
is a sufficient condition that the Frobenius norm of $\Sigma$ is moderate. As in section~2, we find that the balance condition in terms of $||R||_F$ and $||Q||_F$, is simple in simple cases, but delicate in general.

\subsection{The case of the SIR filter}\label{sec:SIRFilter}
The collapse of the SIR filter has been studied in \cite{Bickel,Bickel2,BickelBootstrap}, and it was shown that, for a properly normalized model and data equation, this collapse is governed by the Frobenius norm of the covariance of $Hx^n$; undoing the scaling, and noting that $x^{n-1}$ has covariance $P$ (the steady state solution of the Riccati equation), we find that the Frobenius norm of 
\begin{linenomath*}\begin{equation*}
	\Sigma =H\left(Q+APA^T\right)H^TR^{-1}.
\end{equation*}\end{linenomath*}
governs the collapse of SIR filters. If $||\Sigma||_F$ is moderate, the SIR filter can work even if $m$ or $k$ are large. This condition induces a balance condition for the covariance matrices of the noises which must be satisfied or else the SIR filter fails. For the simple example considered earlier ($A=H=I_m$, $Q=qI_m$, $R=rI_m$), this condition becomes
\begin{linenomath*}\begin{equation*}
	\frac{\sqrt{q^2+4qr}+q}{2r}\leq\frac{1}{\sqrt{m}}.
\end{equation*}\end{linenomath*}
For $m=100$, the $(q,r)$-region for which data assimilation with an SIR filter can be successful is plotted in the right panel of figure \ref{fig:CompareMap}. We observe that this region is very small compared to the region for which data assimilation is feasible with an optimal particle filter. 

We can also set $\epsilon = q/r$ and obtain
\begin{linenomath*}\begin{equation*}
	\frac{\sqrt{\epsilon^2+4\epsilon}+\epsilon}{2}\leq\frac{1}{\sqrt{m}},
\end{equation*}\end{linenomath*}
which we solve for $m$ so that we can plot the maximum dimension for which SIR particle filtering can be successful as a function of the covariance ratio $\epsilon$ (see figure \ref{fig:FilterMap}). Again, we observe that the SIR particle can only be useful in a limited class of problems. In particular, we find that the SIR particle filter works in high-dimensional problems only if the model is very accurate (compared to the data). However, we argued before that this case is somewhat unrealistic, since we expect that the errors in the model be typically larger than the errors in the data (or else the data are not very useful, or particle filtering unnecessary because the model is very good). In these realistic scenarios, the SIR particle filter collapses and we conclude that, as the dimension $m$ increases, it becomes more and more important to use the optimal importance function or a good approximation of it (see e.g. \cite{Morzfeld2011,Brad, Weare2009, Weare 2012} for approximations of the optimal filter).

In the general case, we can use an upper bound, e.g.
\begin{linenomath*}\begin{equation*}
	||\Sigma||_F \leq||H||_F^2||R^{-1}||_F\left(||Q||_F+||A||_F^2||P||\right),
\end{equation*}\end{linenomath*}
and if we require that this bound is less than $\sqrt{m}$, we obtain the simplified balance condition
\begin{linenomath*}\begin{equation*}
	||H||_F^2\left(||Q||_F+||A||_F^2||P||\right)\leq ||R||_F.
\end{equation*}\end{linenomath*}
The above condition implies that the Frobenius norm of the covariance matrix of the model noise, $Q$, must be much smaller than the Frobenius norm of the covariance matrix of the errors in the data, which is unrealistic.

\subsection{Discussion}
We wish to point out differences and similarities of our work and the asymptotic studies in \cite{Bickel,Bickel2,BickelBootstrap,Snyder}. Clearly, the results of \cite{Bickel,Bickel2,BickelBootstrap,Snyder} are used in our analysis of the optimal particle filter (section \ref{sec:OptimalFilter}) and the SIR filter (section \ref{sec:SIRFilter}). Moreover, our analysis confirms Snyder's findings in \cite{Snyder}, that the optimal particle filter is more robust in applications with large $m$ and $k$ because it ``dramatically reduces the required sample size'' (by lowering the exponent in the relation between the number of particles and the state dimension). In \cite{Bickel,Bickel2,BickelBootstrap,Snyder}, it was shown that the number of particles required grows exponentially with the variance of the logarithm of the weights; the variance of the logarithm of the weights is governed by the Forbenius norms of covariance matrices (which are different for SIR and the optimal particle filter). Our main contribution is to study the connection of these Frobenius norms with the effective dimension of section \ref{sec:EffectiveDimension}: if the effective dimension is moderate, then these Frobenius norms can be small even if $m$ or $k$ are large. Thus, one can find conditions under which the SIR and optimal particle filters work. We also explain the physical interpretation of our results and conclude that the optimal/implicit particle filter can work for many realistic and large dimensional problems.

\section{Particle smoothing and variational data assimilation}\label{sec:VarAndSmoothing}
We now consider the role of the effective dimension in particle smoothing and variational data assimilation. The idea here is to replace the step-by-step construction of the conditional pdf in a particle filter (or Kalman filter) by direct sampling of the full pdf 
$p(x^{0:n}|z^{1:n})$, i.e. all available data are assimilated in one sweep. Particle smoothers apply importance sampling to obtain weighted samples from this pdf, and in variational data assimilation one estimates the state of the system by the mode of this pdf.

It is clear that either method can only be successful if the Frobenius norm of the covariance matrix of the variables conditioned on the data is moderate (even if $m$ or $k$ are large), or else the samples of numerical or physical experiments collect on a thin shell far from the most likely state (to obtain this result, one has to repeat the steps in section 2). We now determine the conditions under which this Frobenius norm is moderate. As is customary in data assimilation, we distinguish between the ``strong constraint'' and ``weak constraint'' problem.

\subsection{The strong constraint problem}
In the strong constraint problem one considers a ``perfect model'', i.e. the model errors are neglected and we set $Q=0$ (see e.g. \cite{TalagrandCourtier}). Since the initial conditions determine the state trajectory, the goal is to obtain initial conditions that are compatible with the data, i.e. we are interested in the pdf
\begin{linenomath*}
\begin{align*}
	p(x^0|z^{1:n})\propto& \exp\left(-\frac{1}{2}\left(x^0-\mu_0\right)^T\Sigma_0^{-1}\left(x^0-\mu_0\right)\right)\\
	&\times \exp\left(-\frac{1}{2}\displaystyle\sum_{j=1}^n\left(z^j-HA^jx^0\right)^TR^{-1}\left(z^j-HA^jx^0\right)\right).
\end{align*}
\end{linenomath*}
Straightforward calculation shows that this pdf is Gaussian (under our assumptions) and its covariance matrix is
\begin{linenomath*}\begin{equation*}
	\Sigma^{-1} =\Sigma_0^{-1}+\displaystyle\sum_{j=1}^n(A^j)^TH^TR^{-1}HA^j.
\end{equation*}\end{linenomath*}
As explained above, successful data assimilation for the Gaussian model requires that the Frobenius norm of $\Sigma$ is moderate so that the samples collect on a small and low-dimensional ball, close to the most likely state. The condition for successful data assimilation is a moderate $||\Sigma||_F$, which in turn induces a condition between the errors in the prior~(represented by $\Sigma_0$) and the data~(represented by $R$), which can be satisfied even if $m$ and $k$ are large. The situation is analogous to the balance conditions we encountered before in sequential data assimilation.

\begin{figure}[htbp]
\begin{center}
{\includegraphics[width=0.7\textwidth]{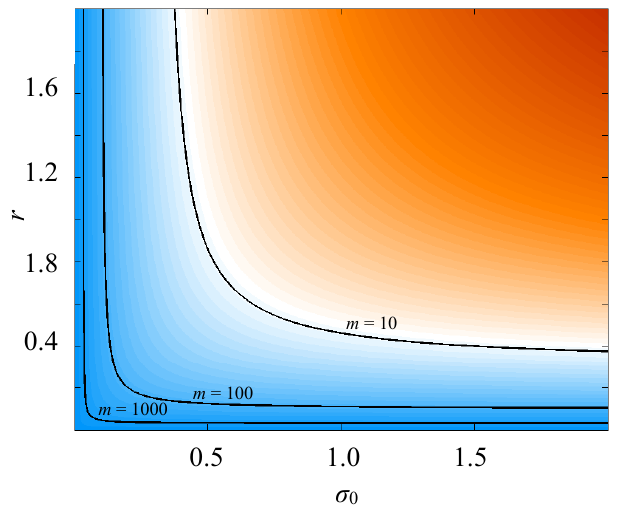}}
\caption{Conditions for successful data assimilation (strong constraint).} 
\label{fig:VarMap} 
\end{center}
\end{figure}

We illustrate the balance condition for the strong constraint problem by considering a version of the simple example we used earlier, i.e. we set $A=H=I_m$, $Q=0$, $R=rI_m$, and, in addition, $n=1$, $\Sigma_0=\sigma_0I_m$. In this case, we can compute $\Sigma$ and its Frobenius norm:
\begin{linenomath*}\begin{equation*}
	||\Sigma||_F =\sqrt{m}\frac{\sigma_0r}{\sigma_0+ r}.
\end{equation*}\end{linenomath*}
Figure \ref{fig:VarMap} shows the values of $r$ and $\sigma_0$ which lead to an $O(1)$ Frobenius norm of $\Sigma$.
Three level sets indicate the state dimensions $m=10,100,1000$; for a given state dimension, the values of $r$ and $\sigma_0$ below the corresponding curve lead to $||\Sigma||_F\approx O(1)$. We observe that, for a fixed $m$, a larger error in the prior knowledge (corresponding to larger values of $\sigma_0$) can be tolerated if the error in the data is very small (corresponding to small values of $r$) and vice versa. Similar observations were made in \cite{Haben2011a,Haben2011b} in connection with the condition number in 3D-Var. Moreover, our analysis confirms what we know from the infinite dimensional problem \cite{Stuart}: as the error in the observation ($r$) goes to zero, the prior ($\sigma_0$) plays no role; however its role is very important even for small observation noise ($r$).

Variational data assimilation (strong 4D-Var) represents the conditional pdf by its mode, i.e. by a single point in the state space. The smaller is the ball on which the samples collect (i.e. the smaller the Frobenius norm of $\Sigma$), the more applicable is strong 4D-Var. Particle smoothers on the other hand construct an empirical estimate of the pdf via sampling. Under our assumptions, we can construct an optimal particle smoother (minimum variance in the weights) by directly sampling the Gaussian posterior pdf (the weights of the particle smoother have zero, thus minimum, variance). We conclude that under realistic conditions (moderate $||\Sigma||_F$) the optimal particle smoother can be expected to perform well, even if $m$ or $k$ are large, because it can efficiently represent the pdf one is interested in. 

The situation is different for other particle smoothers. Consider, for example, the SIR-like particle smoother that uses $p(x_0)$ as its importance function. This filter produces weights whose negative logarithm is given by
\begin{linenomath*}\begin{equation*}
	\phi =\frac{1}{2}\displaystyle\sum_{j=1}^n\left(Z^j-HA^jx^0\right)^TR^{-1}\left(Z^j-HA^jx^0\right).
\end{equation*}\end{linenomath*}
For $n=1$, the variance of these weights depends on the Frobenius norm of the matrix $HA\Sigma_0A^TH^TR^{-1}$, which has the upper bound
\begin{linenomath*}\begin{equation*}
	||HA\Sigma_0A^TH^TR^{-1}||\leq ||H||_F^2||A||_F^2 ||\Sigma_0||_F ||R^{-1}||.
\end{equation*}\end{linenomath*}
If we require that this upper bound is less than $\sqrt{m}$ then we obtain (using $\sqrt{m}\leq||A||_F||A^{-1}||_F$) the condition
\begin{linenomath*}\begin{equation*}
||H||_F^2||A||_F^2 ||\Sigma_0||_F \leq ||R||,
\end{equation*}\end{linenomath*}
which implies that the errors before we collect the data must be smaller than the errors in the data, which is unrealistic. In particular, for the simple example considered above we find that $\sigma_0\leq r/\sqrt{m}$. We conclude that, as in particle filtering, particle smoothing is possible under realistic conditions only if the importance function is chosen carefully.

Note that the results we obtained here are different than those we would obtain if would simply put $Q=0$ in the Kalman filter formulas of section~2. It is easy to show that for $Q=0$ the steady state covariance matrix converges to the zero matrix, provided the dynamics are stable. What this means is that, with enough data, one can wait for steady state, and then accurately estimate the state at large $n$. What we have done in this section is to consider the consequences of having access to only a finite data set, i.e. making predictions before steady state is reached.

Finally, note that, in contrast to the sequential problem, the minimum variance of the weights of the smoothing problem is zero, whereas particle filters always produce non-zero variance weights. This variance is induced by the factorization of the importance function $\pi$, and since this factorization is not required in particle smoothing, this source of variance can disappear (or be reduced) by clever choice of importance functions. As indicated in section \ref{sec:OptimalFilter}, the reason for the reduction in variance of the weights is that the data at time $n$ may render the data at time $n-1$ unlikely; the smoother can make use of this information while the filter can not, since it is ``blind'' towards the future. However, as the data sets get larger (and one eventually runs out of memory), one will have to assimilate the data in more than one sweep, thus inducing additional variance. Ultimately, smoothing as many data sets at a time as feasible can not be a (complete) solution to the data assimilation problem.

\subsection{The weak constraint problem}
In the weak constraint problem (see e.g. \cite{Bennet1993}), one is interested in estimating the full state trajectory given the data, i.e. in the pdf
\begin{linenomath*}
\begin{align*}
	p(x^{0:n}|z^{1:n})\propto& \exp\left(-\frac{1}{2}\left(x^0-\mu_0\right)^T\Sigma_0^{-1}\left(x^0-\mu_0\right)\right)\\
	&\times \exp\left(-\frac{1}{2}\displaystyle\sum_{i=1}^n\left(x^i-Ax^{i-1}\right)^TQ^{-1}\left(x^i-Ax^{i-1}\right)\right)\\
	&\times \exp\left(-\frac{1}{2}\displaystyle\sum_{j=1}^n\left(z^j-Hx^j\right)^TR^{-1}\left(z^j-Hx^j\right)\right).
\end{align*}
\end{linenomath*}
An easy calculation reveals that this pdf is Gaussian and its covariance matrix is 
\begin{linenomath*}\begin{equation*}
	\Sigma^{-1}=\scriptsize\left( \begin{array}{cccc}
\Sigma_0^{-1}+A^TQ^{-1}A & -A^TQ^{-1}  &\cdots & 0\\
-Q^{-1}A 		                     &Q^{-1}+A^TQ^{-1}A+H^TR^{-1}H & -A^TQ^{-1} &        \\
0 					     & \ddots                       &     \ddots  & \ddots    \\
\vdots 					     &                       &            &-A^TQ^{-1}\\
0 					     & \cdots                        &         -Q^{-1}A &Q^{-1}+H^TR^{-1}H \\
\end{array} \right).
\end{equation*}\end{linenomath*}
For the same arguments as before, successful data assimilation requires that the Frobenius norm of $\Sigma$ is moderate. This condition implies (again) a delicate balance condition between the errors in the prior knowledge ($||\Sigma_0||_F$), the errors in the model (\ref{model}) ($||Q||_F$) and the errors in the data (\ref{data}) ($||R||_F$). If this condition is satisfied, data assimilation is possible even if $m$ or $k$ are large.

As in the strong constraint problem, variational data assimilation (weak 4D-Var) represents the conditional pdf by its mode (a single point) and this approximation is the more applicable, the smaller the Frobenius norm of $\Sigma$ is. An optimal particle smoother can be constructed for this problem by sampling directly (zero variance weights) the Gaussian conditional pdf. For the same reasons as in the previous section, we can expect an optimal particle smoother to perform well under realistic conditions, but also can expect difficulties if the choice of importance function is poor.

\section{Limitations of the analysis}\label{sec:limitation}

We wish to point out limitations of the analysis above. To find the conditions for successful data assimilation, we study the conditional pdf and we rely on the Kalman formalism to compute it. Since the Kalman formalism is only applicable to linear Gaussian problems, our results are at best indicative of the general nonlinear\slash non-Gaussian case. However, we believe that the general idea that the probability mass must concentrate on a  low-dimensional manifold  holds in the nonlinear case as well. Since Khinchin's theorem is independent of our linearity assumption, and since we expect that correlations amongst the errors also occur in nonlinear models, one can speculate that the probability mass does collect on a low-dimensional manifold (under realistic assumptions on the noise). However finding (or describing) this manifold in general becomes difficult and is perhaps best done on a case-by-case basis, so that special features of the model at hand can be exploited.

We have further assumed that all model parameters, including the covariances of the errors in the model and data equations, are known. If these must be estimated simultaneously (combined parameter and state estimation), then the situation becomes far more difficult, even in the case of a linear model equation (\ref{model}) and data stream (\ref{data}). It seems reasonable that estimating parameters using data at several consecutive time points (as is done implicitly in some versions of  variational data assimilation or particle smoothing)  would help with the parameter estimation  problem and perhaps even with model specification.

Concerning particle filters, we have examined in detail only two choices of importance function, the one   
in SIR, where the samples are chosen independently of the data, and, at
the other extreme, one where the choice of samples depends strongly on the data. There is a 
large literature on importance functions, see \cite{Brad,Doucet, Weare2009,Weare2012,vanLeeuwen,Ades,Chorintupnas,Morzfeld2011,chorin2010}; it is quite
possible that other choices can outperform the optimal/implicit particle filter even 
in the present linear synchronous case once computational costs are
taken into account. In nonlinear problems the optimal particle filter is hard to
implement and the implicit particle filter is suboptimal, so further analysis 
may be needed to see what is optimal in each particular case (see also \cite{Weare2009, Weare2012} for  approximations of the optimal filter).

More broadly, the analysis of particle filters in the present paper  is not robust as assumptions change. For example, if the model noise is multiplicative (i.e. the covariance matrices are state dependent), then our analysis does not hold, not even for the linear case. Moreover, the optimal particle filter becomes very difficult to implement, whereas the SIR filter remains easy to use. Similarly, if model parameters (the elements of $A$ or the covariances $Q$ and $R$) are not known, simultaneous state and parameter estimation using an optimal particle filter becomes difficult, but SIR, again, remains easy to use. While the filters may not collapse in these cases, they may give a poor prediction. The existence of such  important departures is confirmed by the fact that the ensemble Kalman filter in the ``perturbed observations'' implementation \cite{EvensenBook} and the square root filter \cite{Tippet2003} differ substantially in their performance if the effects of nonlinearity are severe \cite{Lei2010}. However, our analysis indicates that, if  (\ref{model}) and (\ref{data}) hold, the ensemble Kalman filter, the Kalman filter and the optimal particle filter are equivalent in the non-collapse region of the optimal filter. 

Similarly, variational data assimilation or particle smoothing can be successful if (\ref{model}) and (\ref{data}) hold. We expect that variational data assimilation and particle smoothing can be successful in the nonlinear case, provided that the probability mass concentrates on a low-dimensional manifold. In particular, particle smoothing has the potential of extending the applicability of Monte Carlo sampling to data assimilation, since the variance of weights due to the sequential problem formulation in particle filters is reduced (the data at time $2$ may label what one thought was likely at time $1$ as unlikely). This statement is perhaps corroborated by the success of variational data assimilation in numerical weather prediction. However, the number of observations that should be assimilated per sweep depends on the various and competing time scales of the problem and, therefore, must be found on a case by case basis.

Finally, it should be pointed out that we assumed throughout the paper that the model and data equations are ``good'', i.e. that the model and data equations are capable of describing the physical situation one is interested in. It seems difficult in theory and practice to study the case where the model and data equations are incompatible with the data one has collected (although this would be more interesting). For example, it is unclear to us what happens if the covariances of the errors in the model and data equations are systematically under- or overestimated, i.e. if the various data assimilation algorithms work with ``wrong'' covariances.

\section{Conclusions}
\label{sec:Conclusions}
We have investigated the conditions under which data assimilation can be successful, according to a criterion motivated by physical considerations, regardless of the algorithm used to do the assimilation.  We quantified these conditions by defining an effective dimension of a Gaussian data assimilation problem and have shown that this effective dimension must be moderate or else one cannot reach reliable conclusions about the process one is modeling, even when the linear model is completely correct. This condition for successful data assimilation induces a balance condition for the errors in the model and data. This balance condition is often satisfied for realistic models, i.e. the effective dimension is moderate, even if the state dimension is large.

The analysis was carried out in the linear synchronous case, where it can be done in some generality; we believe that this analysis captures the main features of the general case, but we have also discussed the limitations of the analysis. 

Building on the results in \cite{Bickel,Bickel2,BickelBootstrap,Snyder}, we studied the effects of the effective dimension on particle filters in two instances, one in which the importance function is based on the model alone, and one in which it is based on both the model and the data. We have three main conclusions:

\begin{enumerate}
\item The stability (i.e., non-collapse of weights) in particle filtering depends on the effective dimension of the
problem. Particle filters can work well if the effective dimension is moderate even if the true dimension is large (which we expect to happen often in practice).

\item A suitable choice of importance function is essential, or else particle filtering fails even when data assimilation is feasible in principle with a sequential algorithm. 

\item There is a parameter range in which the model noise and the observation noise are roughly comparable, and in which even the optimal particle filter collapses, even under ideal circumstances. 
\end{enumerate}

We have then studied the role of the effective dimension in variational data assimilation and particle smoothing, for both the weak and strong constraint problem. It was found that these methods too require a moderate effective dimension or else no accurate predictions can be expected. Moreover, variational data assimilation or particle smoothing may be applicable in the parameter range where particle filtering fails, because the use of more than one consecutive data set helps reduce the variance which is responsible for the collapse of the filters.

These conclusions are predicated on the linearity of the model and data equations, and on the assumption that the generative and data models are close enough to reality.

\section*{Acknowledgements}
We thank Prof. P. Bickel of UC Berkeley for many interesting discussions, for making our thoughts more rigorous (where possible) and for helping us recognize the limitations of our analysis. We thank Prof. R. Miller of Oregon State University for very helpful discussions and help with the literature. We thank Prof. J. Weare for an interesting discussion. This work was supported in part by the Director, Office of Science, Computational and Technology Research, U.S. Department of Energy under Contract No. DE-AC02-05CH11231, and by the National Science Foundation under grant DMS-1217065.

\bibliographystyle{apalike}
\bibliography{References}

\section*{Figure captions}
Figure 1, Conditions for successful sequential data assimilation.\\
Figure 2, Conditions for successful sequential data assimilation (left panel), and for successful particle filtering; center panel: optimal\slash implicit particle filter; right panel: SIR filter. The broken ellipse in the right panel locates the area where the SIR filter works. \\
Figure 3, Maximum dimension for two particle filters.\\
Figure 4, Conditions for successful data assimilation (strong constraint).
  \end{document}